\documentstyle[aps,prb,epsfig,psfig,twocolumn,overcite]{revtex}
\begin{document}
\title{Refraction at Media with Negative Refractive Index}

\author{ S.~Foteinopoulou$^{1}$, E.~N. Economou$^{2}$, and
C.~M.~Soukoulis$^{1,2,*}$ }

\address{$^1$ Ames Laboratory-USDOE and Department of Physics and
Astronomy,\\ Iowa State University, Ames, IA 50011\\}

\address{$^2$ Research Center of Crete, FORTH, Heraklion, Crete, Greece\\}

\address{$^*$ Author to whom correspondence should be addressed.
Email: soukoulis@ameslab.gov\\}

\maketitle

\abstract
{We show that an electromagnetic (EM) wave undergoes negative
refraction at the interface between a positive and negative
refractive index material.  Finite difference time domain (FDTD)
simulations are used to study the time evolution
of an EM wave as it hits the interface.  The wave is trapped
temporarily at the interface and after a
long time, the wave front moves eventually in the negative direction.
This explains why causality and
speed of light are not violated in spite of the negative refraction
always present in a negative index material.\\
PACS~numbers:~78.20.Ci, 41.20.Jb,42.25.-p, 42.30.-d \\}}

Veselago {\cite{vesalago} predicted that lossless materials, which
possess simultaneously negative permittivity, $\epsilon$, and
negative permeability, $\mu$, would exhibit unusual properties such
as negative index of refraction, $n=-\sqrt{\epsilon \mu}$, antiparallel
   wave vector, $\bf k$, and Poynting vector, $\bf S$, antiparallel
phase, $\bf v_p$, and group, $\bf v_g$, velocities,
and time-averaged energy flux, $\langle{\bf S}\rangle=\langle {u}
\rangle{\bf v_g}$, opposite to the time-averaged momentum density
$\langle{\bf p}\rangle=\langle{u}\rangle{\bf k}/\omega$
\cite{2},  where $\langle {u} \rangle$ is the time-averaged energy
density. Furthermore if these materials are uniform, ${\bf k}, {\bf 
E}, {\bf H}$
form  a left-handed set of vectors. Therefore, these materials are
called left-handed materials (LHM) or negative index of
refraction materials (NIM). The quantities, $\bf S$, u, $\bf p$,
refer to the composite system consisting of EM field
and material. As a result of $\bf k$ and $\bf S$ being antiparallel,
the refraction of an EM wave at the interface
between a positive $n$ and a negative $n$ material would be at the
``wrong'' side relative to the normal (negative refraction).
In addition, the optical length, $\int n d\it l$, is negative in a LHM.

Following Pendry's suggestions \cite{pendry,perfectlens} for specific
structures which can have both $\epsilon_{eff}$ and
$\mu_{eff}$ negative (over a range of frequencies), there have been
numerous theoretical and experimental studies
\cite{natolhm,padilla,aplucsd,science}. In particular, Markos and
Soukoulis \cite{markos} have employed the transfer matrix
technique to calculate the transmission and reflection properties  of
the structure suggested by Pendry \cite{pendry}
and realized experimentally by Smith et al. \cite{padilla,aplucsd}.
Subsequently, Smith et al. \cite{effective}  proved that the
data of Ref. [9] can be fitted by length independent and
frequency dependent,
$\epsilon_{eff}$, and $\mu_{eff}$. They found that in a frequency
region both $\epsilon_{eff}$ and $\mu_{eff}$ were negative
with negligible imaginary parts.  In this negative region, $n$ was
found to be unambiguously negative. These unusual results [3-10]
   have raised objections both to the interpretation of the
experimental data and to the realizability of negative refraction
\cite{garcia,texas}.

In this letter we report numerical simulation results, which clarify
some of the controversial issues, especially the
negative refraction considered as violating causality and the speed
of light \cite{texas}. Our numerical
calculations were performed on a well understood realistic system,  which is
essentially inherently lossless, namely
a 2D photonic crystal (PC). The dielectric constant $\epsilon$
is modulated in space and both the permittivity $\epsilon$ and
permeability $ \mu$ are locally positive.
Specifically, the  photonic crystal consists of an hexagonal lattice of
dielectric rods with dielectric constant, $\epsilon$ = 12.96.  The
radius of the dielectric
rods is r = 0.35a, where a is the lattice constant of the system.
A frequency range  exists for which the EM wave
dispersion is ``almost'' isotropic. For that range the
group velocity, $\bf v_g$,
is antiparallel to the crystal momentum $\bf k$\cite{notomi}.
In particular for the frequency of our simulations
the angle between the group velocity and the phase velocity 
$\theta_{pg}$ varies between $176^0$ and $180^0$.
We have studied systematically the equifrequency contours of the 
system \cite{foteino}
and have obtained the effective refractive 
index, $n$, as a function of frequency. The absolute value of $n$,
$|n|=c|{\bf k}|/\omega$, the phase velocity, ${\bf v_p}=(c/|n|){\bf
k_0}$, where
${\bf k_0}={\bf k}/|{\bf k}|$, and the group velocity ${\bf v_g}=
\partial \omega/\partial {\bf k}$ are defined the usual way.
Then, it can be easily shown that ${\bf v_g}={\bf v_p}/\alpha$, where
   $\alpha=1+(\omega/n)(dn/d\omega)=
1+d\ln |n|/d\ln \omega$;
given this last relation is natural to identify
the sign of $n$ with the sign of $\alpha$ in order
to associate negative $n$ with antiparallel $ \bf{v_p}$
and $ \bf{v_g}$ \cite{15}. Furthermore, it can be shown that
   $\langle{\bf S}\rangle= \langle{u}\rangle{\bf v_g}$, where the
symbol $\langle \hspace{0.1cm} \rangle $ denotes averaged value over time and
over the unit cell. Thus,
${\bf v_p}$ and ${\bf v_g}$ being  antiparallel leads to almost all
the peculiar and interesting properties associated with LHM.

          Essentially this means that for the PC system a frequency range exists
for which the effective refractive index is negative, frequency dispersion
is almost isotropic and $<\bf S>$ $\cdot \bf k < $ 0, i.e. the PC 
behaves as a left-handed (LH)
system.  Consequently a wave hiting the
PC interface for that frequency will undergo negative refraction for the same
reason a wave undergoes negative refraction when it hits the 
interface of a homogeneous
medium with negative index $n$ (the component of $\bf k$ along the
normal to the interface
reverses direction). In Fig.1 we plot our results for the effective refractive
index $n$ for the PC system versus the dimensionless frequency $\tilde{f}$,
where $\tilde{f}={\omega a/2 \pi c}=a/{\lambda}$ and $\lambda$ is the
wavelength in air.  Notice that the effective refractive index $n$
passes continuously from negative to
positive values as $\tilde{f}$ increases. At this point we would like to alert
the reader about other conditions under which negative refraction can 
occur in the air
PC interface. For example, for some particular $\bf k$, light bends 
``the wrong way'' at the PC
interface as a result of the curvature of the equifrequency surfaces 
\cite{Luo} in spite of 
  $<\bf S> \cdot \bf k$, -- and therefore the effective refractive
index--, being positive. Also in a PC system with $<\bf S> \cdot \bf k>$ 0 
coupling to a higher
order Bragg wave can lead to a negatively refracted beam. We would 
like to stress
that in both the cases mentioned above, the PC is right-handed (RH). 
So the origin 
of the negative refraction in these cases is unrelated
with that occurring on
the interface between a RH and LH medium. Thus, these cases should 
not be confused
with the case that we study in this paper where the PC is LH.

In our simulations a finite extent line source was placed outside a
slab of PC at an angle of 30$^{\rm 0}$ as shown in Fig. 2. The source
starts gradually emitting at $t=0$ a monochromatic TE wave (the $\bf E$ field
in the plane of incidence) of dimensionless frequency, $\tilde{f}=0.58$, with
a Gaussian amplitude parallel to the line source. Employing a 
FDTD\cite{Yee,taflove,Beren}
technique, we follow the time and space evolution of the emitted EM waves as
they reach the surface of the PC and they propagate eventually within
the PC. The real space is discretized in a fine rectangular
grid, (of a/31 and a/54 for the x and y axes, respectively),
that stores the dielectric constant, and the electric and magnetic
field values.  By use of a finite time step, $\delta t$=0.0128a/c,
the fields are recursively updated on every grid point.  This
algorithm
numerically reproduces the propagation of the
electromagnetic field in real space and time through the interface. In
Fig. 2 we present the results for our FDTD simulation after 200
simulation steps. (The period, $T=2\pi/\omega$ corresponds to 135 time
steps,
$T \sim 135 \delta t$).

We found, as expected for a negative index $n$, that the incident
beam is eventually refracted in the negative direction
($\sin{\theta_{PC}} \le 0)$. However, the most interesting finding is
that each ray does not refract in the final direction
immediately upon hitting the surface of the PC. Instead, the whole
wavefront is trapped in the surface region for a relatively long time
(of the order of a few tens of the wave period, $T$, in our
simulations); and then, gradually after this transient time, the wave
reorganizes itself and starts propagating in the negative direction
as expected from the steady state solution. Thus, the interface
between the vacuum (positive
$n=1$) and the PC (negative $n$) acts as a strong resonance
scattering center which traps temporarily the wave before gradually
re-emitting it. This time delay (which is much longer that the time
difference, $2t_{0} \simeq 400 \delta t$, between the arrival at the
interface of the inner and outer rays) explains satisfactorily the
apparent paradox of the outer ray propagating much faster than the
velocity of light
\cite{texas}.

         In Fig. 3, we present a time sequence of the amplitude of the
magnetic field of the Gaussian beam undergoing
reflection and refraction at the interface between vacuum ($n=1$) and
a negative index ($n=-0.7$) material. We would like to stress that despite
the fact that the fronts are obscured by the Bloch modulation the refraction
of the wave (after a transient time) is clearly in the negative direction. We have also 
calculated the Poynting
vector (Fig. 4) which shows that (in the steady state) the energy 
flows in the same negative
direction as shown in Fig. 3e. Since one can prove in general that
$\langle{\bf S}\rangle=\langle {u}\rangle{\bf v_g}$, it is clear that 
the group velocity in the PC is
along the direction indicated in Fig. 3e.
  We present the results in terms of the time difference,
$2t_0$, between the arrival of the outer and the inner rays at the interface.
Figure 3a shows the results for t = 2.5$t_{0} \simeq 3.7T \simeq
500\delta t $.  Notice that no refracted front has developed yet.
Fig. 3b
shows the results for t =4.5 $t_0$.  Notice that the wave front has
crossed the interface and seems to move along positive angles.
Figure 3c shows the results for t =10.5 $t_0$, where both the
reflected and refracted wave fronts are shown.  Notice there seem to appear
two fronts for the refracted beam, one moving towards positive angles
and one towards negative angles.  Figure 3d clearly shows that the
wavefront at t = 15$t_0$ along positive angles has diminished and the
wavefront along the negative angles is more pronounced.  Also,
notice the clear reflected wavefront.  Finally, Fig. 3e for t =31
$t_0$ clearly shows that the wavefront moves along negative angles, a
behavior that continues for $t\gtrsim31t_0$.  This is an unambiguous
proof that negative refraction exists in media with a negative
refraction index.  This time sequence of the wavefronts shows that
once the one end of the Gaussian wavefront hits the interface, it
does not mean that the other side of the wavefront must move in``zero
time'' or ``at infinite speed'' to refract
negatively (see Fig. 1b of Ref. 12).  There is a transient time of
the order of 30$t_{0}\simeq 45T$ needed for the refracted wave to
reorganize and move eventually along  the negative directions. Notice
that the wavefronts move finally along negative angles. The same is true
for the directions perpendicular to the wavefronts, which can be
considered  parallel to the group velocity.

We would like to note that Ziolkowski and Heyman\cite{ziol} have observed
a similar transient time effect
numerically in a 1D model system with homogeneous negative permittivity
$\epsilon$
and permeability $\mu$. In [20] it was shown that there
exists a time-lag between the wave hitting the interface and the medium
responding with negative effective index $n$ indicating that causality is
maintained. In the present paper we have shown that in a realistic structure
transient effect precedes the establishment of steady state propagation
in the negative refracted direction \cite{parazzoli}.

We note in passing that our time-dependent results are unrelated with
the steady state solution obtained
   by Valanju et al. \cite{texas}.
  In Ref. 12, a modulated plane wave, composed
of two plane waves with different frequencies,
   is incident on the interface with a certain angle. Indeed, this steady state
solution of propagation of
a modulated plane wave, shows the {\it{interference}} fronts refracting in
the positive direction, while the {\it{phase}} fronts refract in the
  negative direction.  However Smith and Pendry \cite{pendrynew} have
demonstrated that when two frequencies are superimposed in the NIM
with wavevectors that are not parallel - which is the case in [12]-
propagation is {\it crabwise} meaning that the interference fronts are not
the group fronts and should not be associated with the direction of
propagation. In addition, Smith \cite{smithnew} has considered
an incident modulated beam with a finite Gaussian profile. The steady
state solution shows that the group velocity
moves along negative angles, despite the distortion
in the modulated fronts. Negative refraction is also obtained by Lu et. al.
\cite{lu} where they consider more than two frequency components for the
incident wave.

In conclusion, we have shown by a FDTD simulation that an EM wave
coming from a positive $n$ region and hitting a plane interface of a
negative $n$ material refracts eventually in the negative angle
direction. Our detailed time-dependent sequence shows that the wave is
trapped initially at the interface, gradually reorganizes itself,
and finally propagates along the negative angle direction. This
transient time explains why negative refraction does not violate
causality or light speed.
 
We would like to thank P. Marko\v{s} for useful discussions. Ames
Laboratory is operated for the U.S. Department of Energy by Iowa
State
University under Contract No. W-7405-Eng-82. This work was supported
by the director for Energy Research, Office of Basic Energy
Sciences. This work was also supported by DARPA, EU, and NATO grants.

\begin{figure}[b]
\epsfig{file=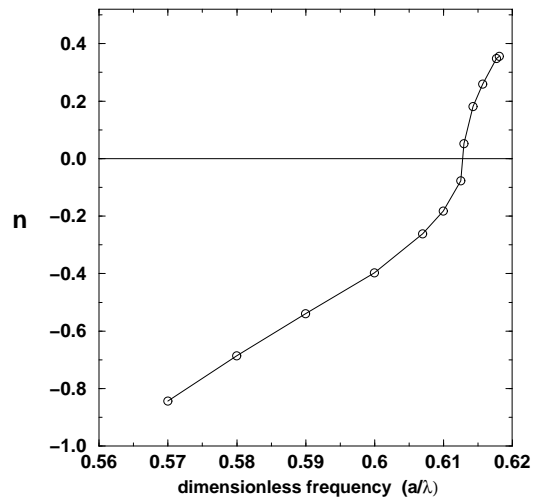,angle=0,width=6.9cm}
\vspace{3mm}
\caption{\label{Figure 1} The effective index of refraction, $n$, 
versus dimensionless
frequency $a/\lambda$ for a 2d photonic crystal.}
\end{figure}

\begin{figure}[t]
\epsfig{file=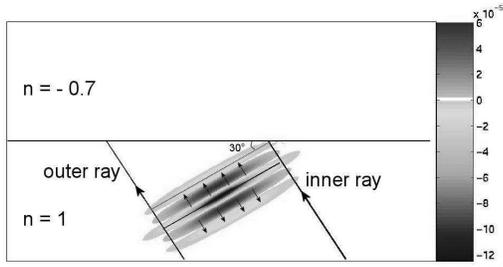,angle=270,width=6.9cm} 
\vspace{3mm}
\caption{\label{Figure 2}
An incident em wave is propagating along a 30$^{\rm 0}$
direction. The time is 200 simulation steps.}
\end{figure}

\begin{figure}
\noindent
\epsfig{file=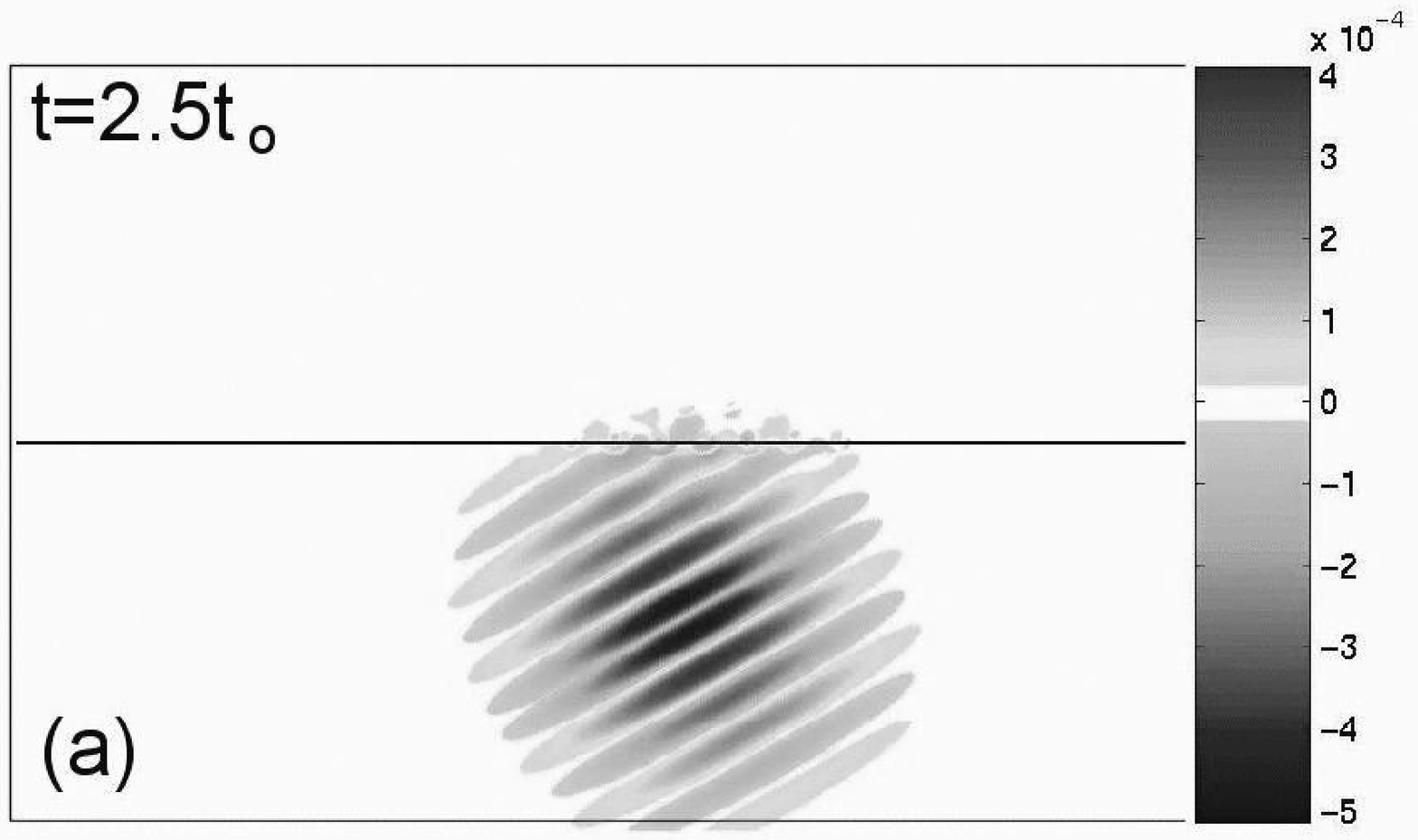,angle=270,width=6.0cm}
\epsfig{file=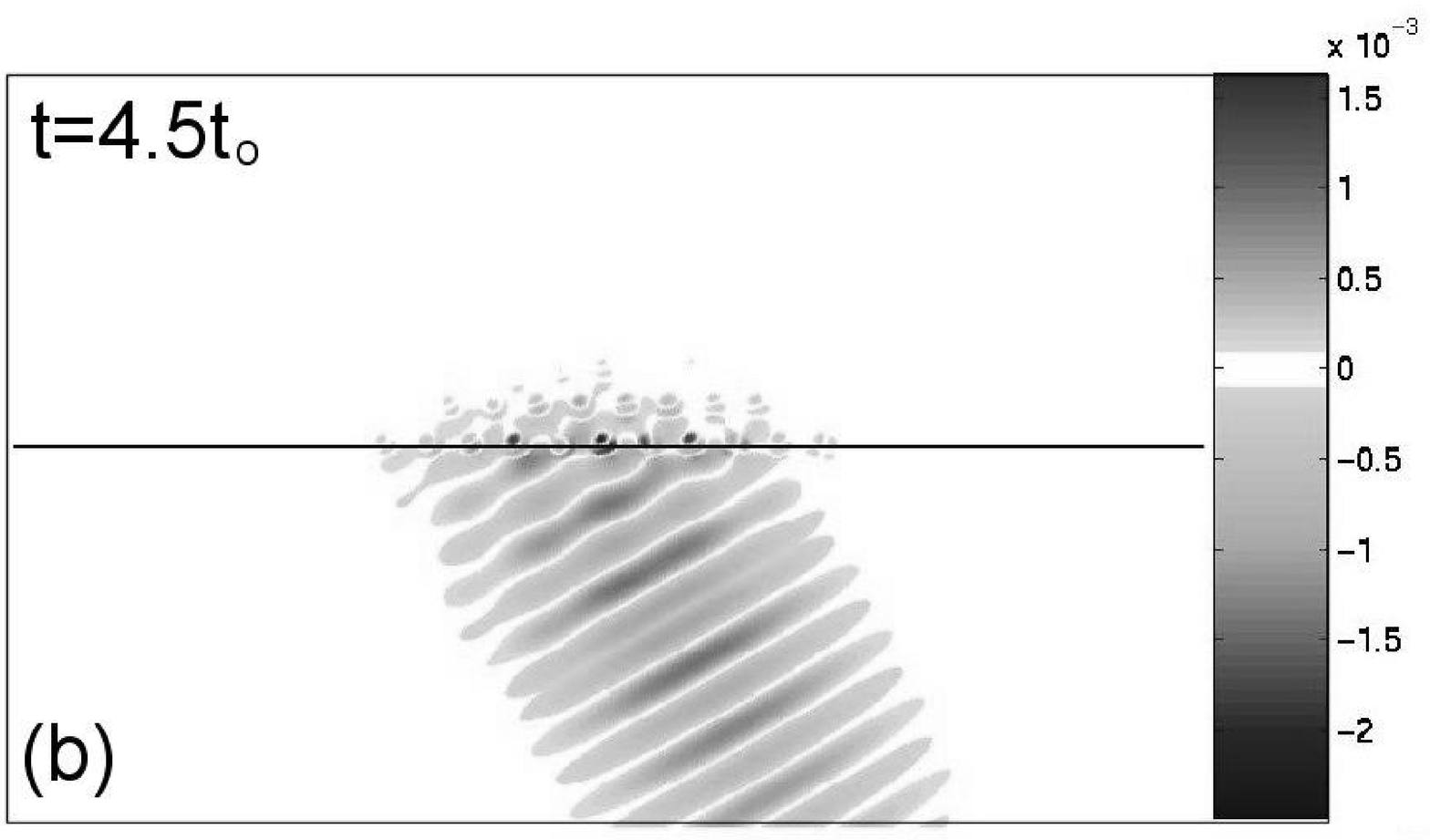,angle=270,width=6.0cm}
\epsfig{file=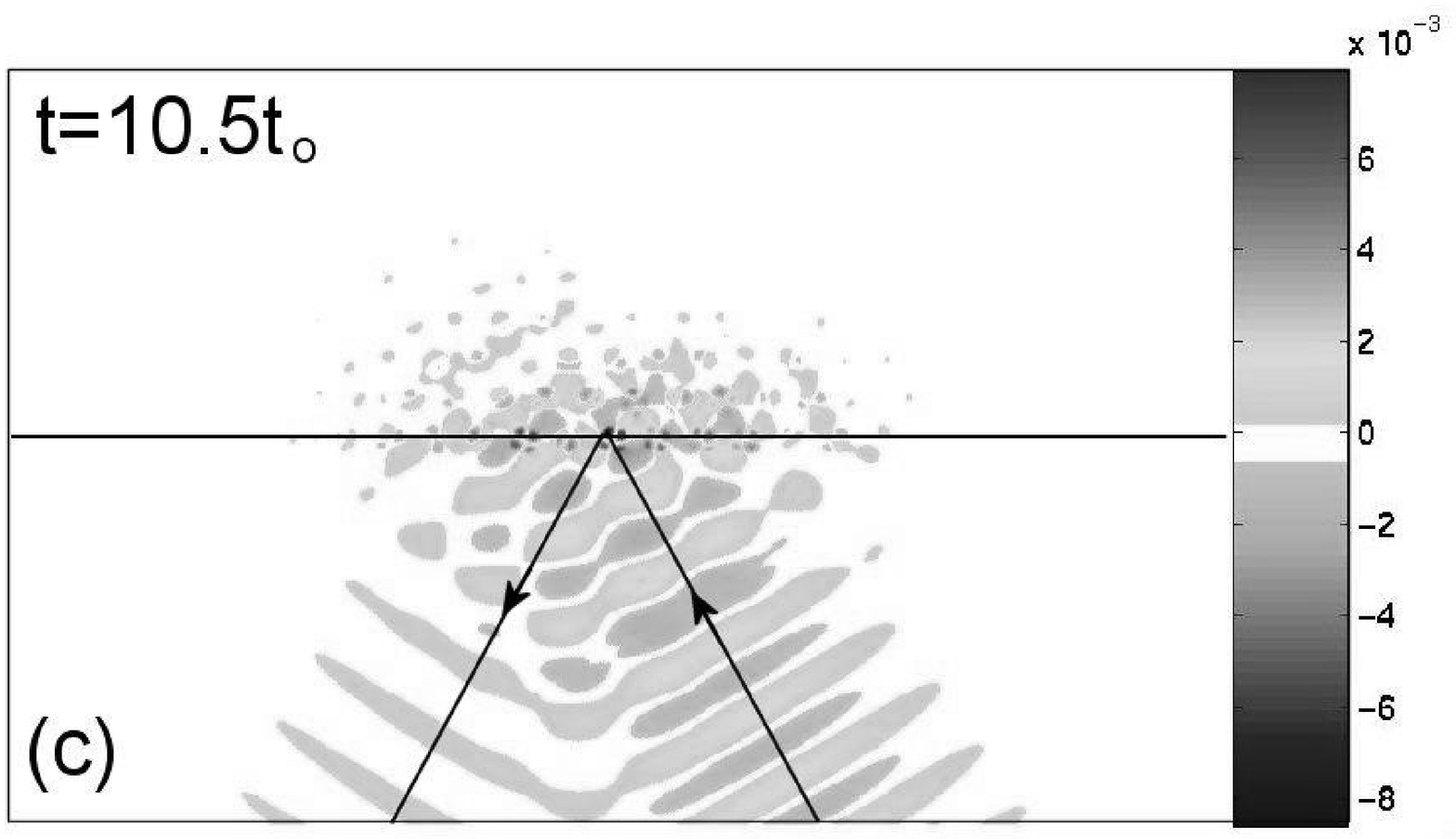,angle=270,width=6.0cm}
\epsfig{file=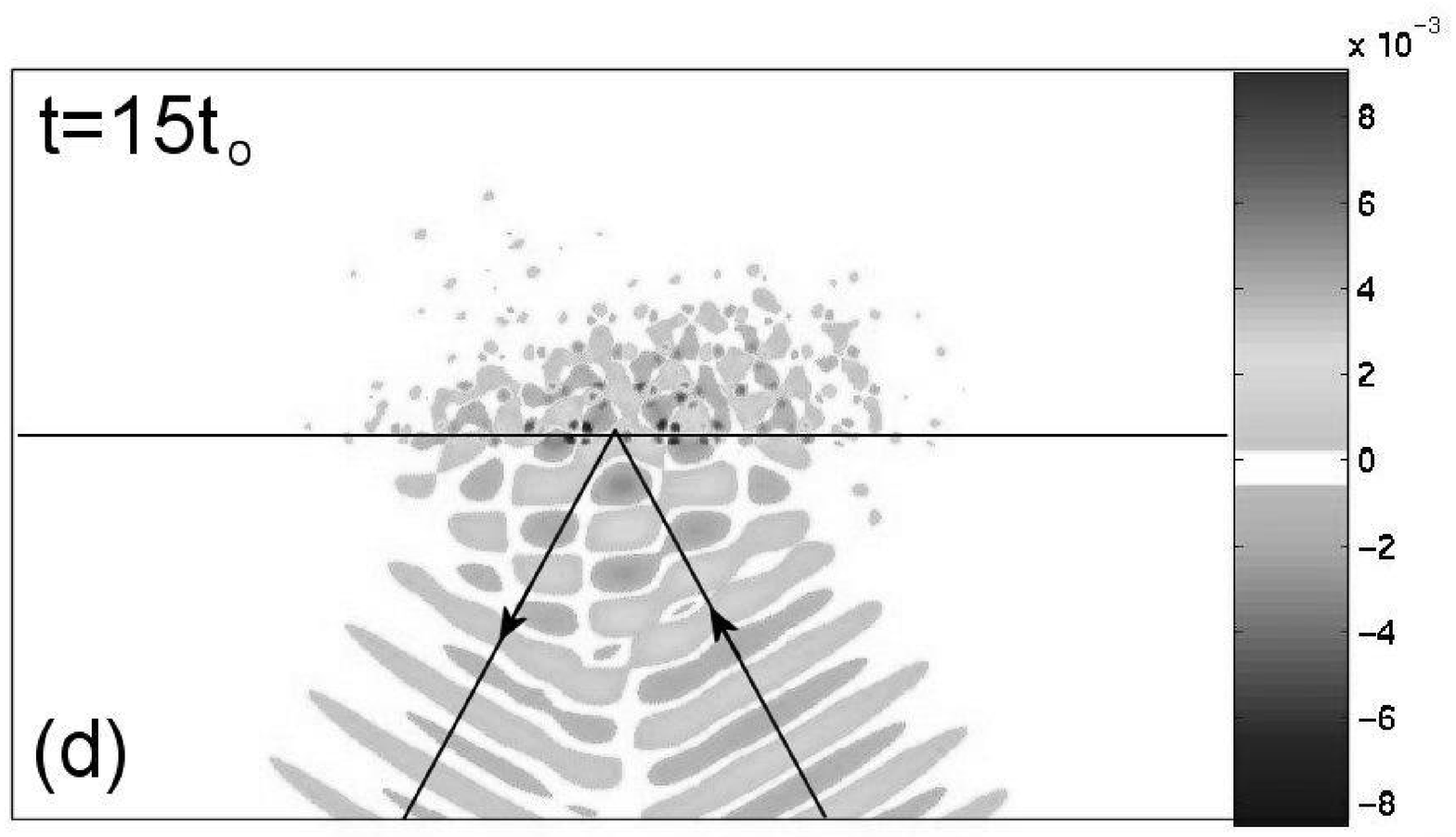,angle=270,width=6.0cm}
\epsfig{file=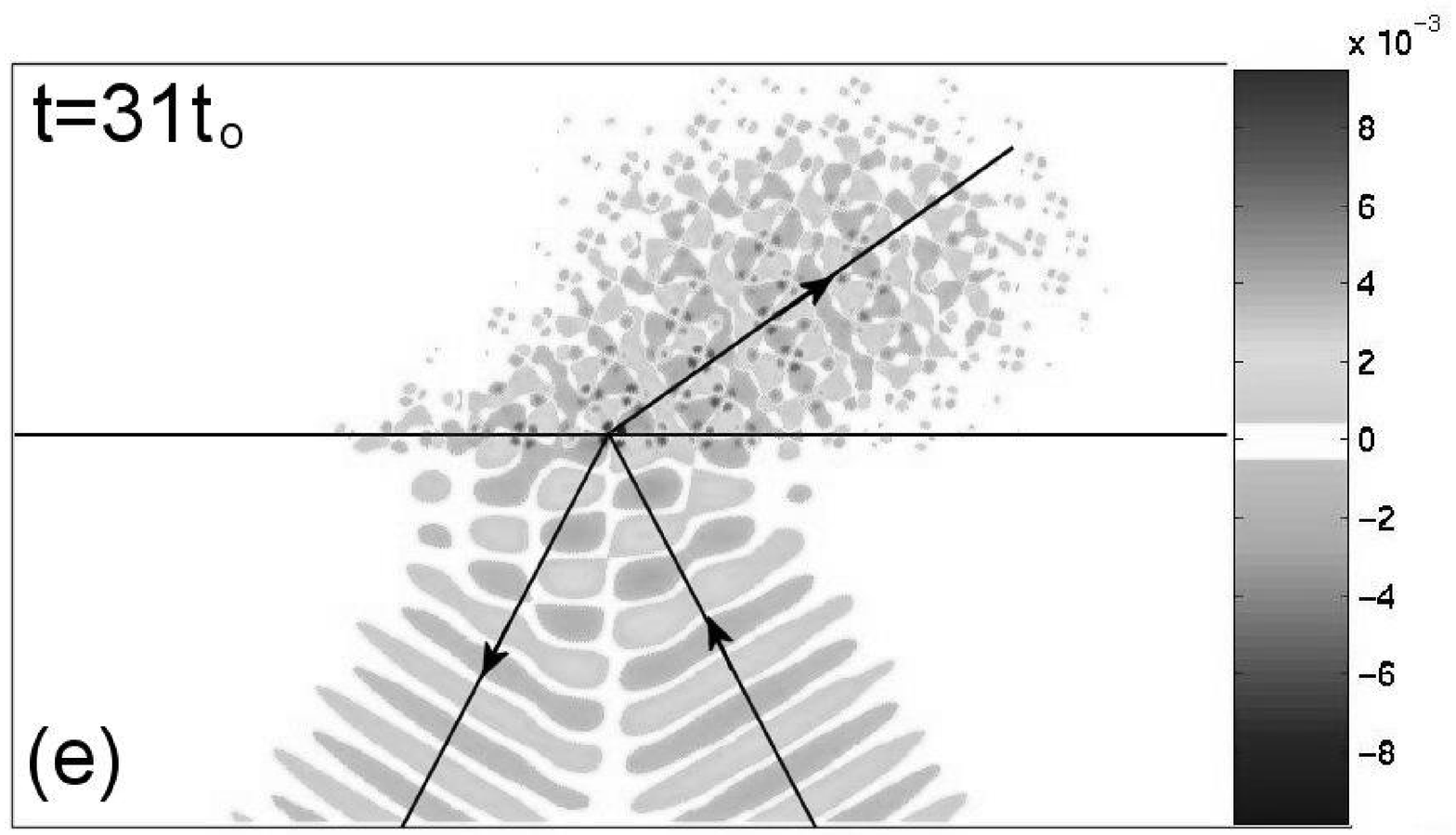,angle=270,width=6.0cm}
\vspace{3mm}
\caption{\label{Figure 3} The magnetic field of the Gaussian beam
undergoing reflection and refraction for (a) $t=2.5t_0$, (b)
$t=4.5t_0$, (c)
$t=10.5t_0$, (d) $t=15t_0$, and (e) $t=31t_0$. $2t_0$ is the time
difference between the outer and the inner rays to reach the
interface;
$t_{0} \simeq 1.5T$, where T is the period $2\pi/\omega$.}
\end{figure}

\begin{figure}[b]
\epsfig{file=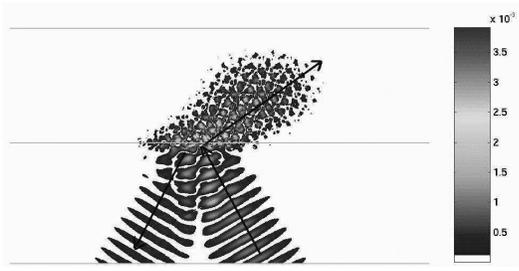,angle=270,width=6.9cm}
\vspace{3mm}
\caption{\label{Figure 4}
The magnitude of the Poynting vector for an EM wave propagating along 
a 30$^{\rm 0}$
direction. The time is 6200 simulation steps, and is the same as the 
one shown in Fig. 3e}
\end{figure}


\begin{thebibliography}{s1}
\bibitem{15} The choice of the sign of $n$ we have adopted here,
i.e., to have $n\alpha \ge 0$, is what one would obtain in a uniform
lossless system where the definition of $n$ is $n=\sqrt{\epsilon\mu}$
if $\epsilon$ and $\mu \ge 0$ and
$n=-\sqrt{\epsilon\mu}$ if $\epsilon$ and $\mu \le 0$. Using this
definition and Kramers-Kronig relations for $\epsilon$ and $\mu$, one
can show that
$\alpha=1+d\ln|n|/d\ln\omega=(1/2)[1+d\ln|\epsilon|/d\ln\omega]+(1/2)[
1+d\ln|\mu|/d\ln\omega]$
has the same sign as $n$. Notice that the $n$ is usually called phase
index of refraction and is
denoted by $n_p$.  If we introduce the group index of refraction
$|n_g|=n\alpha$ we can write that
$|{\bf{v_g}}|=c/|n_g|$.
\end{thebibliography}
\end{document}